\documentclass[baaa]{baaa}

\usepackage[pdftex]{hyperref}
\usepackage{subfigure}
\usepackage{natbib}
\usepackage{helvet,soul}
\usepackage[font=small]{caption}

\contriblanguage{0}

\contribtype{1}

\thematicarea{11}

\title{La orientaci\'on de las iglesias jesu\'iticas en Am\'erica: resultados preliminares}

\titlerunning{La orientaci\'on de las iglesias jesu\'iticas en Am\'erica}

\author{
A. Di Paolo\inst{1}, M.F. Muratore\inst{2,3} \& A. Gangui\inst{2,4}
}

\authorrunning{Di Paolo et al.}

\contact{algangui@gmail.com}

\institute{
Departamento de F\'isica, Facultad de Ciencias Exactas y Naturales, UBA, Argentina \and
Instituto de Astronom\'ia y F\'isica del Espacio, CONICET--UBA, Argentina          \and
Departamento de Ciencias Básicas, UNLu, Argentina                                  \and
Facultad de Ciencias Exactas y Naturales, UBA, Argentina
}

\resumen{ Estudiamos las iglesias misioneras jesuíticas en América, que durante casi dos siglos fueron las construcciones más representativas en el proceso de evangelización cristiana en el continente hasta la expulsión de la Orden en 1767. El objetivo principal es discernir posibles patrones de orientación en las estructuras estudiadas y evaluar si estas orientaciones se relacionan con la salida del Sol o de otros cuerpos celestes por el horizonte local, como sugieren los textos de los primeros escritores cristianos, lo que podría brindar información importante y novedosa acerca de su historia y construcción. En el pasado se han realizado trabajos de campo arqueoastronómicos en dos extensas regiones de Sudamérica: las provincias históricas de Paraquaria y de la Chiquitanía (oriente de la actual Bolivia). Estos datos deben ser ahora interpretados dentro de un contexto cultural y geográfico más amplio. Así, para obtener una imagen más completa de la arquitectura religiosa en el continente, es necesario analizar y comparar los resultados con la orientación de las iglesias jesuíticas edificadas en los siglos XVII y XVIII en América del Norte (virreinato de Nueva España). Abordamos el proyecto a través del análisis histórico y cultural de los antiguos sitios misionales y por medio de imágenes satelitales. En este trabajo presentamos algunos resultados preliminares a los que hemos llegado.}

\abstract{ We study the Jesuit mission churches in America, which for almost two centuries were the most representative constructions in the process of Christian evangelisation on the continent until the expulsion of the Order in 1767. The main objective is to discern possible orientation patterns in the structures studied and to evaluate whether these orientations are related to the rising of the sun or other celestial bodies on the local horizon, as suggested by the texts of early Christian writers, which could provide important and novel information about their history and construction. Archaeoastronomical fieldwork has been carried out in the past in two large regions of South America: the historic provinces of Paraquaria and Chiquitanía (eastern present-day Bolivia). These data must now be interpreted within a broader cultural and geographical context. Thus, in order to obtain a more complete picture of religious architecture on the continent, it is necessary to analyse and compare the results with the orientation of Jesuit churches built in the 17th and 18th centuries in North America (viceroyalty of New Spain). We approached this project through the historical and cultural analysis of the former mission sites and by means of satellite imagery. In this paper we present some preliminary results we have reached.}


\begin{document}

\maketitle

\section{Introducción}\label{S_intro}

El estudio de la disposición de las iglesias cristianas ha sido de interés durante mucho tiempo, y ha recibido un nuevo impulso en la literatura reciente al reconocerse que la orientación
espacial de sus ejes principales representa un rasgo clave de su arquitectura. Según los textos de los primeros escritores cristianos, los altares de las iglesias debían estar en una
dirección particular, es decir, el sacerdote tenía que situarse mirando hacia el oriente durante los oficios. Así lo reconocen Orígenes, Clemente de Alejandría y Tertuliano, y habría sido
ratificado en el primer Concilio de Nicea. También en el siglo IV San Atanasio de Alejandría declaró que el sacerdote y los participantes debían mirar hacia el este, de donde Cristo, el
Sol de Justicia, brillaría al final de los tiempos \citep{mccluskey-2015}. 

Si bien se han realizado estudios históricos y culturales detallados de los pueblos misioneros sudamericanos y de sus edificios más emblemáticos, la orientación de las iglesias (o de sus
ruinas) en estos pueblos no había sido objeto de un estudio en profundidad hasta hace poco tiempo. En los últimos años se llevó a cabo el trabajo de campo arqueoastronómico que considera
las características urbanísticas y los escritos y crónicas de los misioneros en los territorios de los pueblos guaraníes, provincia de \emph{Paraquaria} \citep{gimenezbenitez-etal-2018}. El trabajo de campo también se extendió recientemente a un estudio arqueoastronómico de las orientaciones de las iglesias jesuíticas en \emph{Chiquitos}, hoy la región de la Chiquitanía en el oriente de Bolivia \citep{gangui-2020}. Pero aparte de estos grupos misionales, las iglesias jesuíticas en el resto de América no han sido debidamente estudiadas con el fin de hallar sus orientaciones espaciales y las razones culturales subyacentes.

En las siguientes secciones introducimos el contexto cultural y describimos los resultados que se han obtenido en diferentes regiones del continente.

\section{La Compañía de Jesús}

Como respuesta de la Iglesia Católica a la Reforma Protestante, la orden de los jesuitas (miembros de la Compañía de Jesús) fue reconocida como una nueva orden religiosa por el papa Pablo
III en 1540. Fundada por el sacerdote vasco Ignacio de Loyola, ``la Compañía'' se convirtió en la guía espiritual e intelectual de la Europa católica durante los dos siglos y medio
siguientes. A través de árduos y extensos viajes en el extranjero, su misión era difundir el cristianismo y salvar almas, propagar la fe en tierras no convertidas. En el interim,
establecieron una red mundial cohesionada que atrajo a personas prominentes y ricas. El predominio de los jesuitas aportó poder a la organización e influencia en todos los ámbitos de la
vida, así como favores políticos, contribuciones financieras e increíbles libertades. Los gobernantes de los Habsburgo de España los favorecieron porque se ofrecían como voluntarios para
los retos más difíciles y cumplían sus promesas, ganándose la desconfianza y la envidia de sus competidores religiosos. Una de esas duras tareas, por supuesto, era la evangelización de
nativos en las más recientes tierras conquistadas por la Corona, por ejemplo, en el Nuevo Mundo. Este favoritismo empezó a decaer cuando los gobernantes borbónicos de España llegaron al
poder en el 1700, con su estricto control de las finanzas y sus sospechas hacia las actividades y operaciones comerciales de los jesuitas. Poco más de medio siglo más tarde la Compañía
sería expulsada de América y luego disuelta.

\section{Los Jesuitas en Nueva España}

Hacia 1572, pocas décadas después de su fundación, los sacerdotes de la Compañía llegaron y se establecieron en Centroamérica, al igual que lo hicieron los miembros de las órdenes
mendicantes de los dominicos y franciscanos. En la zona de interés para nuestro estudio, a partir de aproximadamente 1640 comienzan a establecerse misiones permanentes en el desierto de Sonora y, más tarde, en 1697, se funda el primer puesto de la península de Baja California que sobrevivió hasta hoy, la misión de Nuestra Señora de Loreto Conchó. Con el tiempo, Loreto se convirtió en la cabecera de las misiones de la península y extendió su influencia en las regiones cercanas. Desde esos sitios estratégicos, tanto en la península como en el continente, los jesuitas atraían a los nativos americanos nómadas y los agrupaban en las misiones para bautizarlos y convertirlos. Durante más de cien años organizaron a los indios para que vivieran en sociedad y cultivaran la tierra. Y estos últimos fueron una asistencia fundamental en la construcción de las sólidas y vistosas iglesias de piedra que poblaron la región. 

\section{Iglesias: antecedentes en Sudamérica}

A partir de los estudios en Sudamérica pudo demostrarse que, a diferencia de las iglesias de Paraquaria donde las orientaciones son mayoritariamente meridianas en dirección norte-sur, la mitad de las iglesias chiquitanas estudiadas tienen orientaciones canónicas que parecen estar alineadas con fenómenos solares, con sus altares orientados hacia declinaciones astronómicas que caen dentro del rango solar y con tres iglesias notables que exhiben una orientación equinoccial precisa \citep{gangui-2020}.
En el proyecto, a la luz de nuestros trabajos anteriores, tratamos de comprender si fue el nuevo paisaje de la selva virgen sudamericana lo que hizo que los misioneros se apartaran de las
indicaciones de los antiguos escritores católicos (como ya sucede, en parte, con las iglesias coloniales en algunas islas de Canarias, e.g., \cite{di_paolo_la_gomera_2020}) o si, por el
contrario, hubo una intención inicial en la costumbre jesuita de orientar sus iglesias monumentales (su modo particular de construir, o \emph{``modo nostro''}, \cite{garofalo-2015}) que podría justificar los datos que fueron obtenidos en estos sitios.
Está bien documentada la adaptabilidad de los jesuitas a las diversas circunstancias, y el perfecto equilibrio que se les recomendaba para la configuración de todas las actividades o políticas de la Compañía, y esto podría también incluir el diseño y la construcción de sus iglesias.


\section{Las iglesias jesuíticas en Nueva España}

Los resultados de los estudios arqueoastronómicos ya desarrollados en América del Sur deben ser interpretados dentro de un contexto histórico y geográfico más amplio. Porque sabemos que antes de las iglesias estudiadas en Chiquitos, por ejemplo, los jesuitas construyeron durante décadas una gran cantidad de iglesias al otro lado del ecuador (e.g., en Sonora y en estados vecinos del México actual). Por lo tanto, para obtener una imagen más completa de la arquitectura religiosa en el continente, es necesario comparar los datos ya analizados con la orientación de las iglesias jesuíticas de los siglos XVII y XVIII en el antiguo virreinato de Nueva España en Norteamérica. 

\begin{figure}[!t]
\centering
\includegraphics[width=\columnwidth]{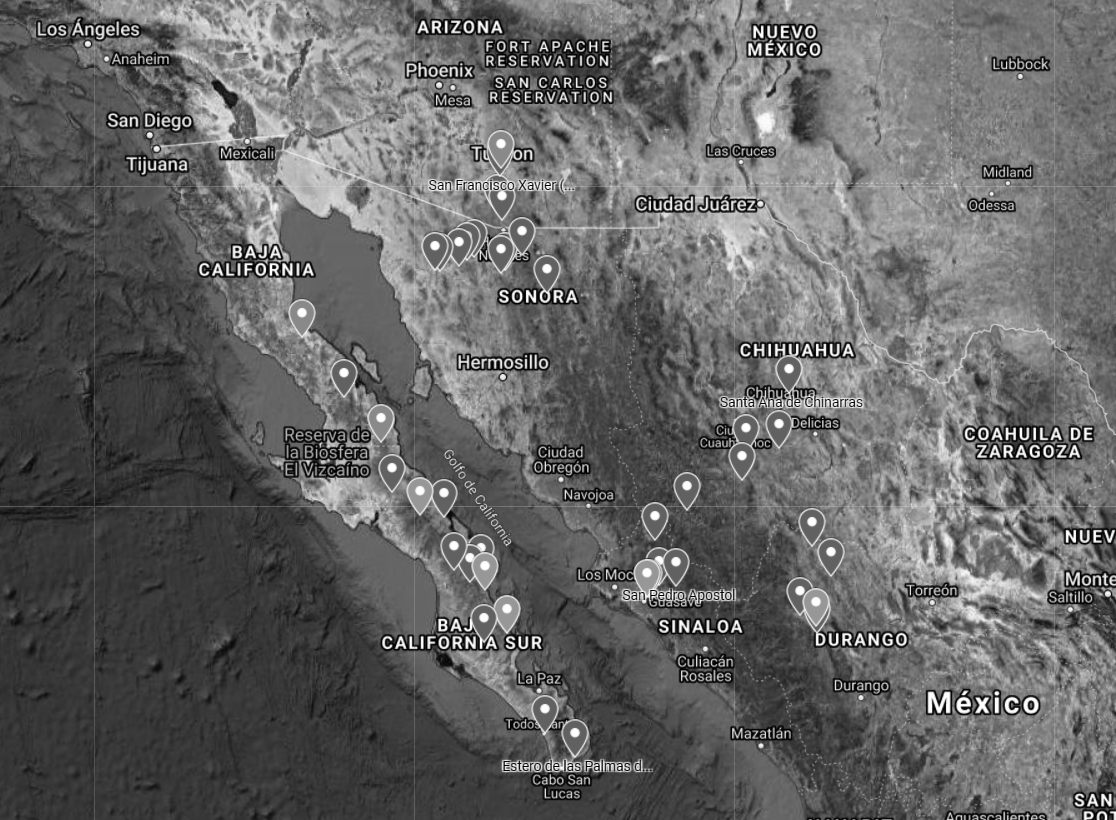}
\caption{Grupo de iglesias jesuíticas de los siglos XVII y XVIII con medidas de acimut. Se ubican en territorios que pertenecieron al virreinato de Nueva España.}
\label{Figura1}
\end{figure}

\section{Metodología y selección de datos}

Realizamos una exploración detallada de los actuales estados mexicanos de Baja California, Sonora, Chihuahua, Sinaloa, Durango, y de otras regiones cercanas, en busca de iglesias
jesuíticas aún en pie o de ruinas de antiguos templos debidamente documentados y con una estructura distinguible. Utilizamos mapas satelitales junto a diversas fuentes documentales y
seleccionamos un número grande de iglesias existentes en la zona que fueron construídas por la Compañía en los siglos XVII y XVIII y para las cuales fuese posible la medición de sus
orientaciones espaciales (Fig. 1). Mediante el empleo de los mapas y de las herramientas de medición de Google Earth podemos estimar, con una precisión de algunos grados, los
acimuts reales de los ejes principales de las iglesias.

\section{Discusión y trabajo en curso}

En la Fig. 2 se presenta el diagrama de orientación para las iglesias analizadas. Las líneas diagonales del gráfico señalan los acimuts correspondientes -en el cuadrante oriental- a los
valores extremos para el Sol (acimuts de $63.0^\mathrm{o}$ y $116.3^\mathrm{o}$ -líneas continuas-, equivalente a los solsticios de verano e invierno boreales, respectivamente) y para la
Luna (acimuts: $57.0^\mathrm{o}$ y $123.2^\mathrm{o}$ -líneas rayadas-, equivalente a la posición de los lunasticios mayores).

\begin{figure}[!t]
\centering
\includegraphics[width=\columnwidth]{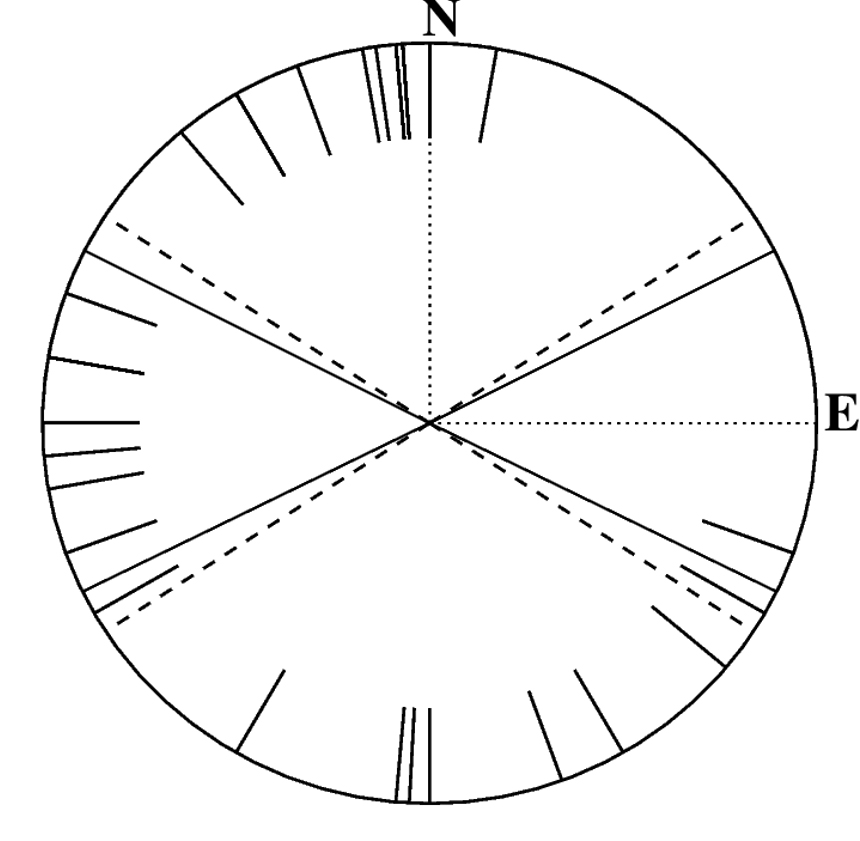}
\caption{Diagrama de orientación preliminar de las iglesias jesuíticas de Nueva España.}
\label{Figura2}
\end{figure}

Nuestros resultados preliminares muestran que una gran mayoría de las iglesias que permanecen en pie en la región histórica de Nueva España se orientan hacia los cuadrantes norte y occidental. Aunque el estudio aún no está completo, sorprende que por el momento muy pocas iglesias muestran orientaciones dentro del rango solar hacia el cuadrante oriental. Esto difiere de otros grupos de templos cristianos coloniales (incluso jesuíticos, como algunos de orientación equinoccial que fueron medidos en Chiquitos) y requiere de una explicación. Lamentablemente, las fuentes hasta ahora consultadas que describen algunos aspectos de las construcciones jesuíticas no ponen el énfasis en la orientación espacial de sus iglesias históricas.

Por otra parte, sabemos que un análisis completo de nuestros datos requiere tomar en cuenta la altura angular del horizonte ``h'' en la dirección de los altares de las iglesias. Estas
medidas pueden obtenerse \emph{in situ} o, en su defecto, y con buena precisión, mediante el empleo de modelos topográficos digitales de elevación, como por ejemplo el que se basa en el
Shuttle Radar Topographic Mission (SRTM) disponible en el sitio heywhatsthat.com/. Un trabajo futuro, aún en progreso, nos permitirá combinar medidas locales de acimut y altura angular
para estimar la declinación astronómica, coordenada que ya no dependerá de la ubicación geográfica ni de la topografía regional. El valor de esta coordenada ecuatorial, calculado para un
dado templo, una vez comparado con la declinación del Sol (que fija aproximadamente un par de días en el año, o sólo uno en el caso de los solsticios), nos permitirá verificar, entre
otras cosas, si esta construcción histórica está o no orientada en una dirección que coincide con la fecha de su fiesta patronal, y evaluar el peso estadístico de estos resultados. Este
análisis, al igual que una investigación exhaustiva de las fechas de construcción de los templos en fuentes documentales, muchas aún desconocidas, está actualmente en desarrollo y
esperamos poder completarlo y reportarlo en otra ocasión. 

\bibliographystyle{baaa}
\small
\bibliography{aaa-dipaolo-muratore-gangui-baaa-2021}
 
\end{document}